\documentclass{jfm}

\usepackage{graphicx}
\usepackage{newtxtext} 
\usepackage{newtxmath} 
\usepackage{natbib}
\usepackage{hyperref}
\hypersetup{
    colorlinks = true,
    urlcolor   = blue,
    citecolor  = black,
}

\usepackage{amsmath} 
\usepackage{bm} 
\usepackage{upgreek} 
\usepackage[version=4]{mhchem}
\usepackage{siunitx}
\usepackage{subcaption} 
\usepackage{float} 
\usepackage{ulem} 
\usepackage{cancel} 
\usepackage{nicefrac}

\newcommand{\bmpsi}{\bm{\uppsi}}
\newcommand{\bmphi}{\bm{\upphi}}

\newcommand{\q}{\mathbf{q}}

\newcommand{\x}{\mathbf{x}}

\newcommand{\Q}{\mathbf{Q}}
\newcommand{\Qhat}{\hat{\mathbf{Q}}}
\newcommand{\W}{\mathbf{W}}
\newcommand{\A}{\mathbf{A}}

\newcommand{\U}{\mathbf{U}}

\newcommand{\POD}{\text{POD}}

\shorttitle{Boundary Layer Transition as Succession of Temporal \& Spatial Symmetry Breaking}
\shortauthor{C. Lin, O.T. Schmidt}

\title{Boundary Layer Transition as Succession of Temporal and Spatial Symmetry Breaking}

\author{Cong Lin\aff{1}, 
Oliver T. Schmidt\aff{1} \corresp{\email{oschmidt@ucsd.edu}}}

\affiliation{\aff{1}Department of Mechanical and Aerospace Engineering, University of California, San Diego, La Jolla, CA 92093, USA} 

\begin{document}
\maketitle
 
\begin{abstract}
We show that both temporal and spatial symmetry breaking in canonical K-type transition arise as organized hydrodynamic structures rather than stochastic fluctuations. Before the skin-friction maximum, the flow is fully described by a periodic, spanwise symmetric, harmonic response to the Tollmien–Schlichting wave, forming a spatially compact coherent structure that produces hairpin packets. This fundamental harmonic response may visually resemble turbulence, but remains fully periodic and delimits the exact extent of the deterministic regime.
A distinct regime change occurs after this point: a hierarchy of new (quasi-)periodic and aperiodic space-time structures emerges, followed shortly by anti-symmetric structures that develop similarly despite no anti-symmetric inputs, marking the onset of aperiodicity and spanwise asymmetry. We identify these structures as symmetry-decomposed spectral and space-time proper orthogonal modes that resolve the full progression from deterministic to broadband dynamics. The key insight is that laminar-turbulent transition can be viewed as a sequence of symmetry breaking events, each driven by energetically dominant, space-time coherent modes that gradually turn an initially harmonic flow into broadband turbulence.

\end{abstract}



\vspace{-44pt}
\section{Introduction}\label{sec:introduction}




Boundary layer transition through deterministic input actuation has been extensively studied to understand fundamental mechanisms underlying the path to turbulence. \citet{fasel1990numerical}
and \citet{rist1995direct} conducted seminal direct numerical simulations (DNS) of controlled K-type transition in flat-plate boundary layers, successfully reproducing wind tunnel experiments by \citet{klebanoff1962three,kachanov1977nonlinear,kachanov1984resonant}.
\citet{bake2002turbulence} studied, through a comparison of DNS and experiments, the complex flow randomization process that transforms small disturbances into developed boundary layer turbulence, providing local phenomenological insights into the breakdown of organized structures. \citet{sayadi2013direct} demonstrated that complete K- and H-type transitions converge towards fully developed turbulence after the skin-friction overshoot, revealing that initially periodic ``hairpin packets'' eventually produce statistical properties of fully developed turbulence. 
Towards the stability analysis of canonical transition, \citet{herbert1988secondary} established a framework for analyzing secondary instability mechanisms in the K- and H-type transition scenarios. \citet{monokrousos2010global}
identified optimal forcing mechanisms and initial conditions for instability growth in the Blasius boundary layer, while \citet{
cherubini2011minimal} investigated optimal perturbations and minimal seeds using time-domain optimization approaches. 
\citet{rigas2021nonlinear} extended input$/$output analysis to account for nonlinear triadic interactions via harmonic balance models, solving transition scenarios up to before the skin friction maximum ($C_f^{max}$) within the frequency domain.

\begin{figure}
  \centering
    \includegraphics[trim={0.0cm 1.5cm 0.4cm 0.25cm },clip,width=1.0\textwidth]{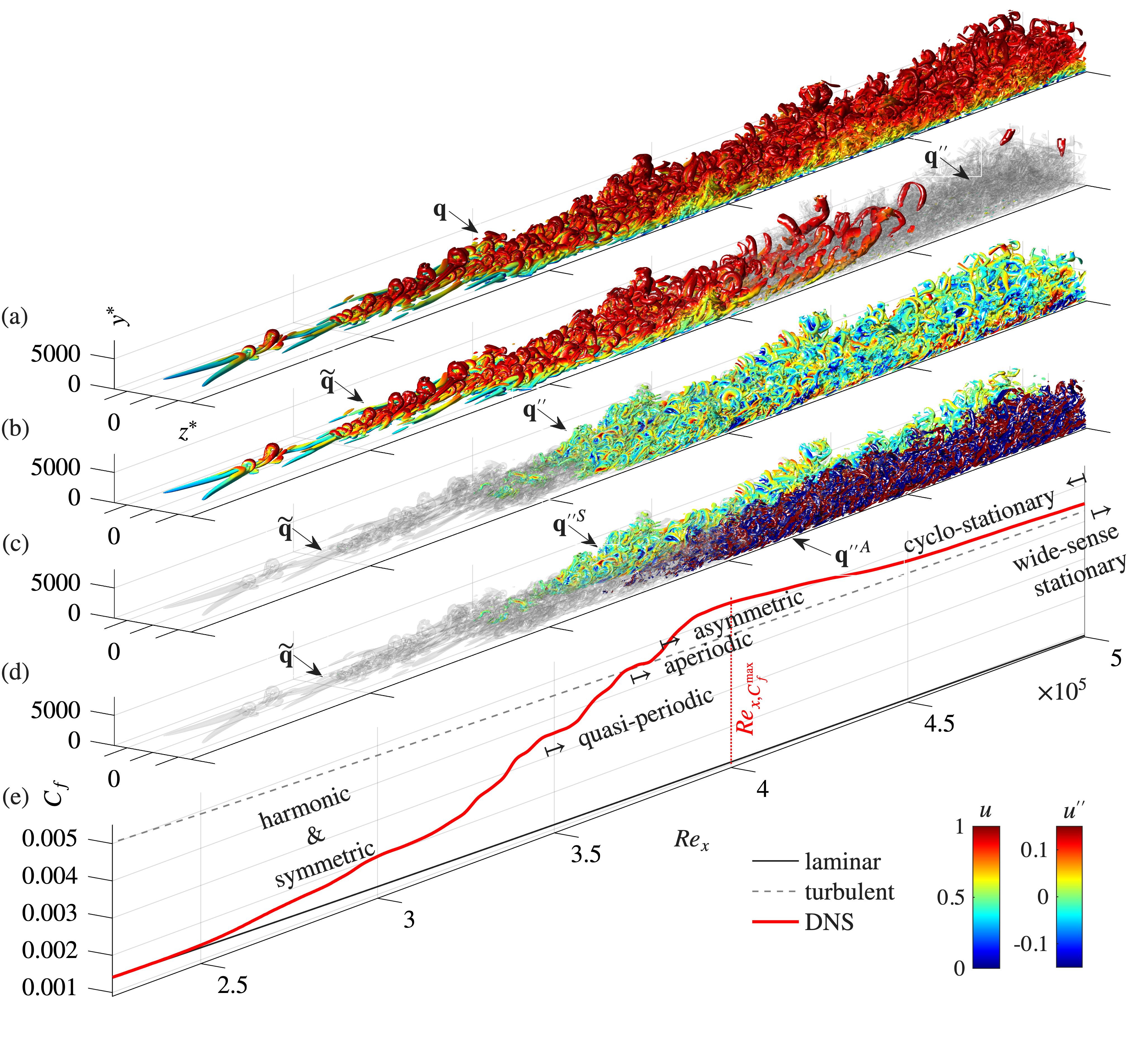}
\vspace{-3pt}    
  \caption{Dynamical regimes of the transition process. (a) Instantaneous DNS snapshot. (b) Fundamental harmonic response $\tilde{\q}$ (STPOD mode $\bmphi_0$). (c) Cyclo-stationary fluctuation $\q''$. (d) Symmetric $\q''^S$ and anti-symmetric $\q''^A$ fluctuation components for time and spatial symmetry breaking. $Q$-criterion isosurfaces ($Q=10^3$ for all); colored in their respective $u$-velocities ($w$-velocity for $\q''^A$). (e) Regimes labeled on the $C_f$ graph.}
  \label{fig:DNS_vs_STPODm1_vs_pcyc}
\vspace{-7pt}    
\end{figure}

Modal analysis has been a valuable tool for identifying coherent structures in transitional flows. \citet{rempfer1994dynamics,rempfer1994evolution} pioneered the application of proper orthogonal decomposition (POD) \citep{sirovich1987turbulence_1,aubry1991hidden} to identify three-dimensional coherent structures and their dynamics in K-type transition. Recent advances in modal decomposition include spectral POD \citep{Lumley1970AP,TowneEtAl2018JFM} for extracting frequency domain modes and space-time POD \citep{schmidt2019conditional,frame2023space}
for general spatio-temporal modes without imposing assumptions on time dynamics. Both methods have been applied to data from fully turbulent boundary layers \citep{tutkun2017lumley,hack2021extreme}.
\citet{heidt2024spectral} extended SPOD to harmonically forced flows, assuming a frequency domain ansatz with specific periodic modal forms.

Despite these advances, the specific mechanisms driving the breakdown of temporal (periodicity) and spatial (spanwise) symmetries---the hallmark phenomena of chaos and turbulence---in deterministic transition remain unclear.
In figure~\ref{fig:DNS_vs_STPODm1_vs_pcyc}, we provide an overview of the streamwise development of flow dynamics and symmetries: The initially periodic and symmetric flow departs from the laminar solution and undergoes temporal and spatial symmetry breaking, marked by increasing non-periodicity and asymmetry, as it progresses through the $C_f^{max}$. Afterwards, the cyclo-stationary state converges to statistical stationarity as the statistics asymptotically collapse onto turbulent correlations \citep{white2006viscous}.
Throughout our analysis, we track the evolution from deterministic periodic states to chaotic and broad-band dynamics, and characterize a variety of coherent structures for these respective regimes.
We demonstrate that both temporal and spatial symmetry breaking are identifiable as organized, energetically dominant modes with distinct dynamical properties.



\vspace{-11pt}
\section{Methods}
We compute the DNS of K-type transition with the \textit{CharLES} solver on a structured mesh, satisfying $y^+ < 1$ across the entire Reynolds number range from $Re_x = 1 \times 10^5$ to $6.5 \times 10^5$. Transition is triggered by a forcing strip that introduces a periodic ($f_1 = 8.75$) Tollmien-Schlichting (TS) wave with a weak symmetric peak ($f_0 = 0$). For details on the numerics and validation, we refer the reader to \cite{lin2024modal}.
To isolate the dominant symmetry breaking mechanisms, a dataset of 198 forcing periods ($T_1=1/f_1$) sampled with 128 DNS snapshots each is decomposed via three methods: D1 symmetry decomposition, Space-Time Proper Orthogonal Decomposition, and Spectral Proper Orthogonal Decomposition. 


\vspace{-5pt}
\subsection{$D_1$ Symmetry Decomposition} \label{subsec:D1}
Leveraging the natural transverse $D_1$ (dihedral group 1) symmetry of the simulation setup and actuation, we decompose the flow data $\q(\x, t)$ into spatially symmetric $\q^S(\x,t)$ and anti-symmetric $\q^A(\x,t)$ components \citep{sirovich1987turbulence_2} with respect to the $z=0$ midplane. 
The symmetric component $\q^S$ of the flow field is defined, with the top set of signs, as
\begin{equation}\label{eq:q_S}
\q^{S/A}(x, y, z ,t) = \frac{1}{2} \left[
\begin{array}{c}
u(x, y, z ,t) \pm u(x, y, -z ,t) \\
v(x, y, z ,t) \pm v(x, y, -z ,t) \\
w(x, y, z ,t) \mp w(x, y, -z ,t)
\end{array}
\right] 
\end{equation}
while the anti-symmetric component $\q^A$ is given by
the bottom set of signs.
Applying this before modal decomposition guarantees modes with pure spatial (anti-)symmetry properties.

\vspace{-5pt}
\subsection{Spectral Proper Orthogonal Decomposition (SPOD)} \label{subsec:SPOD}
SPOD identifies frequency-specific, time-periodic coherent structures $\bmpsi(\bm{x},f)$ in statistically stationary flows 
\citep{Lumley1970AP, TowneEtAl2018JFM, SchmidtColonius2020AIAAJ}. Here, the space-time correlation tensor $\bm{C}(\bm{x},\bm{x}',\tau)$ (where $\tau=t-t'$) is Fourier-transformed to the cross-spectral density (CSD) tensor $\bm{S}(\bm{x}, \bm{x}', f) = \int_{-\infty}^{\infty} \bm{C}(\bm{x}, \bm{x}', \tau) e^{-2\pi i f \tau} d\tau$ to solve the respective continuous and discrete eigenvalue problems
\begin{align}\label{eq:SPOD_EVP}
\int_{V} \bm{S}(\bm{x},\bm{x}', f) \bm{W}(\bm{x}') \bmpsi(\bm{x}', f) \, d\bm{x}' &= \lambda(f) \bmpsi(\bm{x}, f), \\
\nicefrac{1}{N_b} \hspace{2pt} \Qhat_{k}\Qhat_{k}^*\W\bm{\Psi}_{k} &= \bm{\Psi}_{k}\bm{\Lambda}_{k},
\label{eq:SPOD_discrete} 
\end{align}
where $\Qhat_{k}$ stores different realizations of the Fourier modes $\hat{\q}_{k}(\x)$ at frequency $f_k$ and the CSD is estimated as $\textbf{S}_k=E_b\{\hat{\q}_{k}(\x)\hat{\q}^*_{k}(\x) \}$ with block-wise expectation $E_b\{\cdot\}$ \citep{Welch1967IEEE}. The columns of $\bm{\Psi}_{k}$ are discrete SPOD modes $\bmpsi_{k}^{(m)}(\x)$ and $\bm{\Lambda}_{k}=\text{diag}(\bm{\uplambda}_{k})$ stores their energies $\uplambda_{k}^{(m)}$. Expansion coefficients are recovered in the rows of $\A_{k} = \bm{\Psi}_{k}^*\W\Qhat_{k}$, with $\A_{k}^*\A_{k} = \bm{\Lambda}_{k}$.  
SPOD modes are single-frequency oscillatory coherent structures that optimally capture the variance of the data at that frequency and are orthonormal in the space- and infinite-time norm $\langle\cdot, \cdot \rangle_{\bm{x},t}$. 
They are thus statistically optimal linear combinations of Fourier modes.

\vspace{-5pt}
\subsection{Space-Time Proper Orthogonal Decomposition (STPOD)} \label{subsec:STPOD}
STPOD extracts space-time-energy-optimal spatio-temporal modes $\bmphi(\bm{x},t)$ over a finite time window $\Delta T$, chosen here as one actuation period $T_1$. The modes solve the respective continuous and discrete eigenvalue problems \citep{Lumley1970AP, schmidt2019conditional}
\begin{align}\label{eq:FredholmEVP}
\int_{\Delta T} \int_V \bm{C}(\bm{x},\bm{x}',t,t') \bm{W}(\bm{x}') \bmphi(\bm{x}', t') \, d\bm{x}' \, dt' &= \lambda \bmphi(\bm{x}, t), \\
\nicefrac{1}{N_{j}} \hspace{2pt} \Q\Q^*\W\bm{\Phi} &= \bm{\Phi}\bm{\Lambda},
\label{eq:STPOD_discrete}
\end{align}
over a spatial domain $V$, with positive-definite weight matrix $\bm{W}(\bm{x})$ and statistically estimated two-point space-time correlation tensor $\bm{C}(\bm{x},\bm{x}',t,t')=E\{\q(\bm{x}, t)\q^*(\bm{x}', t') \}$. The modes $\bmphi(\bm{x},t)$ are orthonormal in the space- and finite-time norm $\langle\cdot, \cdot \rangle_{\bm{x},\Delta T}$.
For the discrete problem, snapshots of $\q(\x, t)$ are stacked column-wise over $\Delta T$ for each realization $j$, and different realizations are concatenated row-wise to form the data matrix $\Q$, which is then decomposed by (\ref{eq:STPOD_discrete}). 
The discrete STPOD modes $\bmphi_m(\x,t)$ form the columns of $\bm{\Phi}$ 
and $\bm{\Lambda}=\text{diag}(\bm{\uplambda})$ stores their energies $\uplambda_m$.
Realization-varying expansion coefficients $a_{m,j}$ are obtained as the rows of $\A = \bm{\Phi}^*\W\Q$, with $\A^*\A = \bm{\Lambda}$.
Notably, STPOD specializes to space-only POD in the limit $\Delta T \to 0$ and, assuming statistical stationarity, converges to SPOD in the limit $\Delta T \to \infty$ \citep{frame2023space}. Here, 
STPOD is the most unbiased method with no assumptions about the dynamics of the space-time signal $\q(\x, t)$. Unlike spectral approaches that impose periodicity by default, it produces periodic modes only when they are energetically dominant within the given window $\Delta T=T_1$ ---a crucial property for symmetry breaking analysis.

\vspace{-10pt}
\section{Extracting the Resonant Fundamental Harmonic Response (FHR)}\label{sec:3}
\begin{figure}
\hspace{-6pt}
\begin{subfigure}[b]{0.3625\textwidth}
  \includegraphics[width=\textwidth]{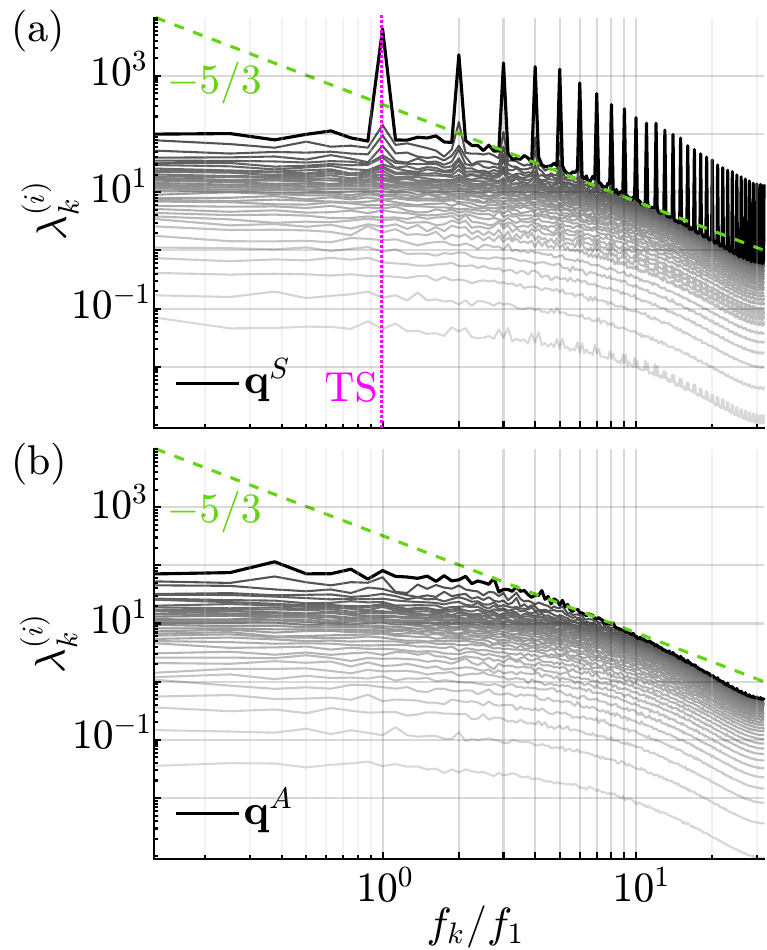}
  \label{subfig:SPOD_lambda_A}
\end{subfigure}
\hfill
\begin{subfigure}[b]{0.635\textwidth}
\centering
  \includegraphics[width=\textwidth]{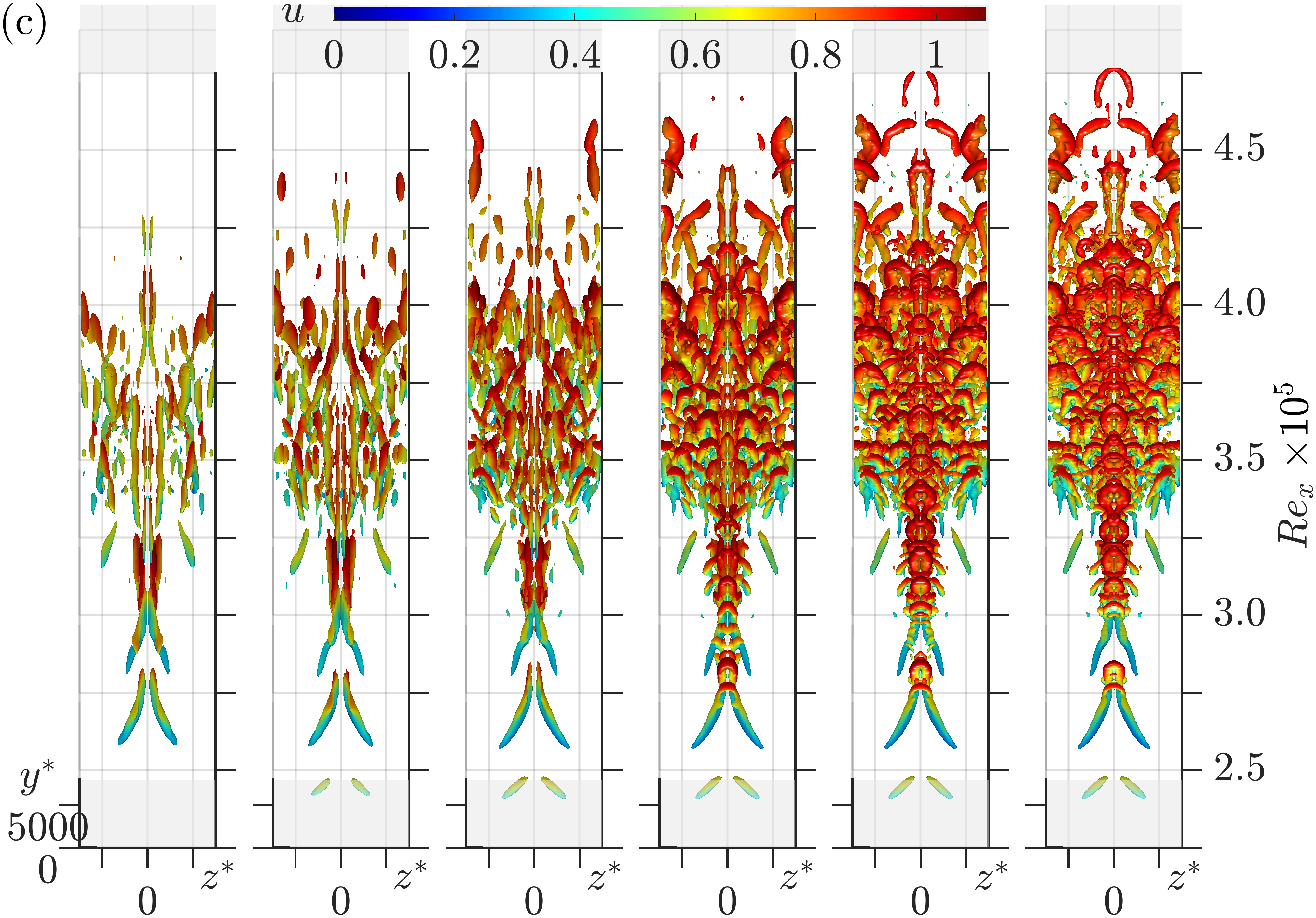}
  \label{subfig:SPOD_lincomb}
\end{subfigure}

\vspace{-12pt}
  \caption{SPOD energy spectra for (a) $\q^S$ symmetric and (b) $\q^A$ anti-sym. components; compare to \citet{sayadi2013direct}. (c) Superposition of $N_n$=\{1, 2, 4, 8, 16, 32\} dominant SPOD modes $\bmpsi^{(1)}_{n}$ at harmonic peaks $f_n$  (\ref{eq:SPOD_lincomb}); $Q$-criterion ($Q=10^3$) colored by $u$.}
  \label{fig:SPOD_panel}
  \vspace{-11pt}
\end{figure}

As we are actuating the eigenmode (TS wave) of a system with quadratic nonlinearities, a resonant response with harmonics must be expected.
To extract those harmonic coherent structures, we apply SPOD to the symmetric ($\q^S$) and anti-symmetric ($\q^A$) components individually, yielding energy-optimal modes at each frequency in each symmetry class.
Figure~\ref{fig:SPOD_panel}(a,b) contrasts the $\q^S$ and $\q^A$ spectra:
The $\q^S$ spectrum is dominated by low-rank peaks at the TS frequency ($f_k/f_1=1$) and its harmonics, characteristic of dominant coherent structures, while the noise-like $\q^A$ spectrum is broadband with slow rank decay, showing no energetically preferred modes.
Thus, the \textit{fundamental harmonic response} (FHR) is symmetric, time-periodic, and represents the expected deterministic result from symmetric, periodic actuation of the Navier-Stokes equations, as predicted by classical frameworks like harmonic balance \citep{rigas2021nonlinear}. 
The FHR can be assembled by a linear combination of the first symmetric SPOD modes 
$\bmpsi^{(1)}_{k}$ of the harmonic peaks $f_k=f_n$ ($n \in \mathbb{Z}$) in figure~\ref{fig:SPOD_panel}(a):
\begin{equation}\label{eq:SPOD_lincomb}
\tilde{\q}(\x,t) \equiv \sum_{n} \bmpsi^{(1)}_{n} a^{(1)}_{n} e^{(i2\pi f_n t)} + \Bar{\q}(\x) 
= 
\tilde{\q}_{\symbol{92} 0}(\x, t) + \Bar{\q}(\x), 
\end{equation}
with the time-averaged mean flow $\Bar{\q}(\x)$, 
and the block-wise estimated dominant SPOD expansion coefficients $a^{(1)}_{n}$.
We compute the SPOD using $N_{\text{FFT}} = 512$, eight periods per block and 50\% block overlap.
Most notably, figure~\ref{fig:SPOD_panel}(c) shows that $\tilde{\q}(\x,t)$ exhibits compact spatial support when fully reconstructed, and, while the resulting vortical structures closely resemble hairpin ``packets''$/$``forests" in its late stages---often characterized as ``chaotic'' in the turbulence literature
\citep{adrian2007hairpin, wu2009direct}---we 
show that such seemingly disordered structures in fact arise \textit{deterministically} here as the system's 
resonant FHR
to periodic actuation of its eigenmode.
As illustrated in figure~\ref{fig:DNS_vs_STPODm1_vs_pcyc}, separating
this deterministic baseline from the full data yields a turbulent, cyclo-stationary fluctuation 
\begin{equation}\label{eq:cyclostat}
\q''(\x, t) \equiv \q(\x, t) - \tilde{\q}(\x, t)= \q''^S(\x, t) + \q''^A(\x, t).
\end{equation}
In case of phase-locked data, we note $\tilde{\q}$ can coincide with the phase mean, although SPOD provides a more general estimate of the harmonics. Further decomposing the fluctuation into symmetry components 
isolates, by construction, the data components containing temporal and spatial symmetry breaking: As $\tilde{\q}$ is purely symmetric and periodic, $\q''^S$ must contain all non-periodic symmetric fluctuations that break time symmetry, while any indication of $\q''^A$ captures the onset of spanwise asymmetry (we note $\q''^A\equiv \q^A$, as this system exhibits no deterministic anti-symmetric flow).
(\ref{eq:cyclostat}) results in a pivotal multiscale separation: 
We later show in figure~\ref{fig:STPOD_energy}(c) that the rapid emergence of $\q''^S$ and then $\q''^A$
comes with a simultaneous decline in the $\tilde{\q}$ amplitude, allowing us to locate exactly where symmetry breaking arises.

\begin{figure}
\centering
\begin{subfigure}[]{0.49\textwidth}
\includegraphics[width=\textwidth]{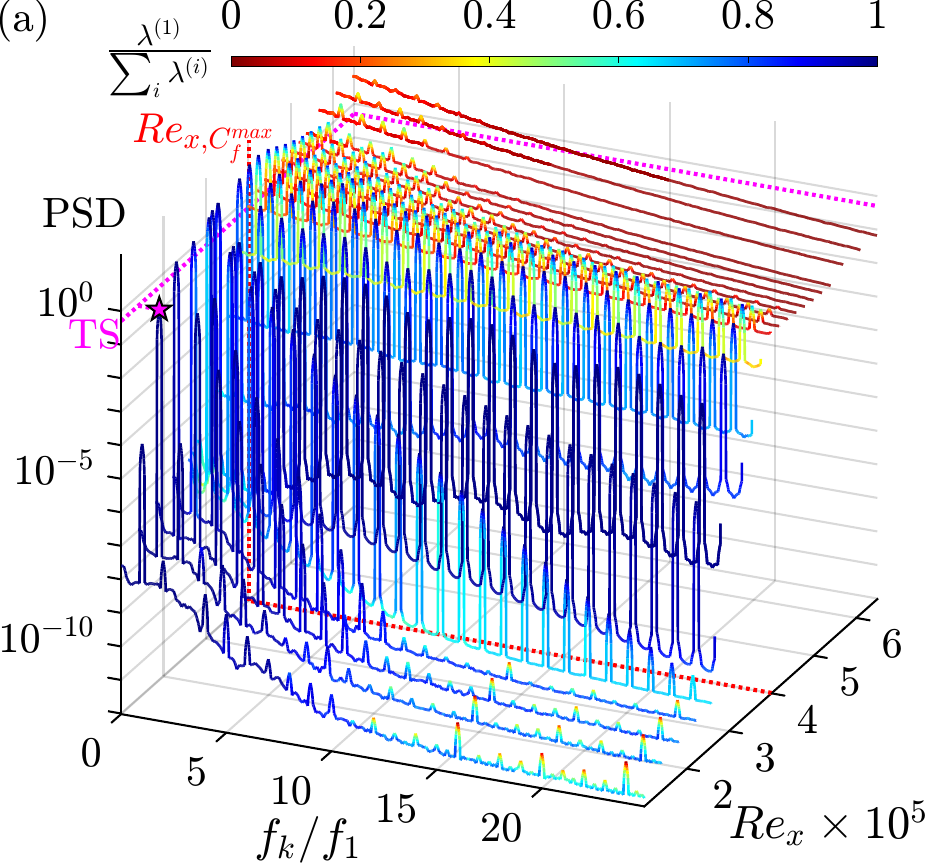}

\end{subfigure}
\hfill
\begin{subfigure}[]{0.49\textwidth}
\includegraphics[width=\textwidth]{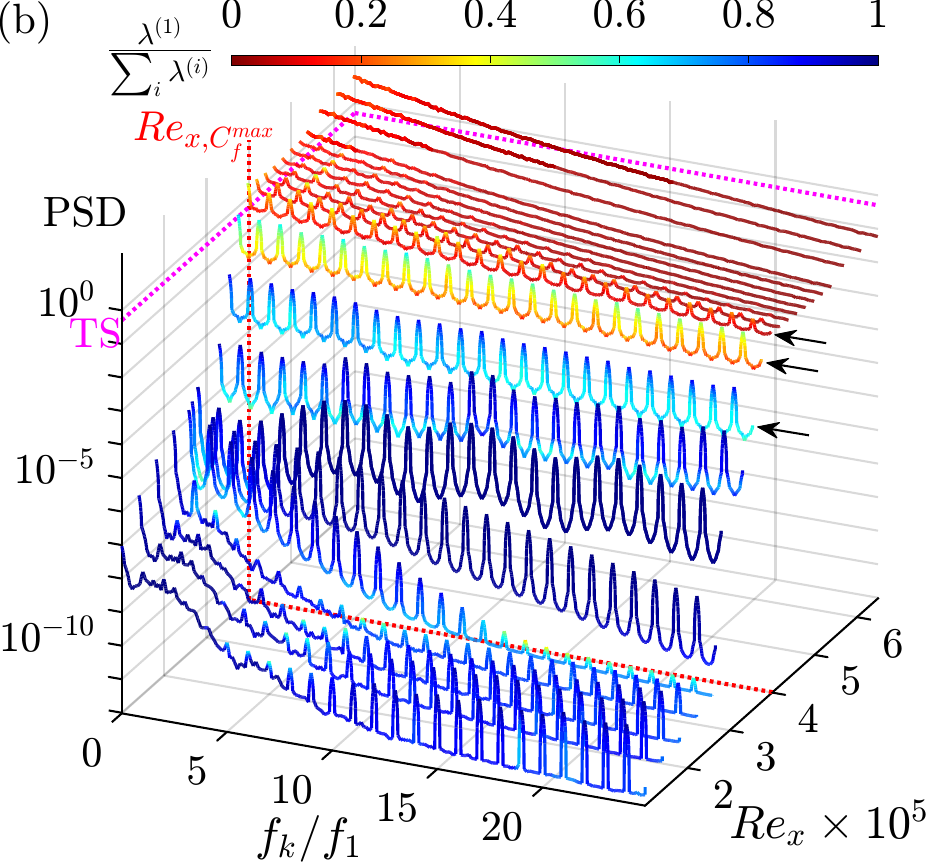}
\end{subfigure}

\vspace{6pt}
\begin{subfigure}[]{0.8\textwidth}
\includegraphics[width=\textwidth]{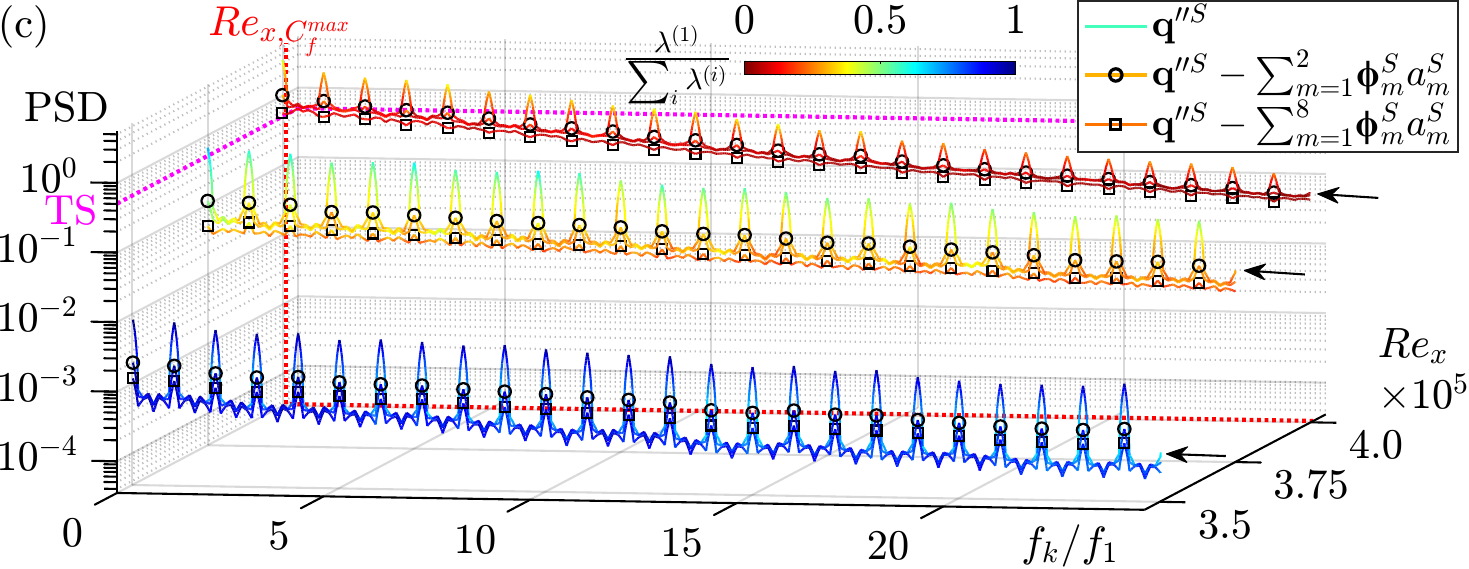} 
\end{subfigure}

\vspace{-3pt}

\caption{Local power spectral densities (PSD) of different quantities as a streamwise function of $Re_x$, colored by the ratio of the dominant local SPOD energy over the full local energy $\uplambda^{(1)}(x)/\sum_{i} \uplambda^{(i)}(x)$. (a) $\q^S(\x,t)$ full symmetric data. 
(b) $\q''^S(\x,t)$ fluctuation data. (c) Zoom on amplified region, $r=\{2,8\}$ first two and first eight symmetric STPOD-mode-subtracted fluctuation $\q''^S(\x,t)-\sum_{m=1}^{r} \bmphi_m^{S}(\x,\uptau) a_{m,j}^S$.}
\label{fig:Lspodx_PSD}
\vspace{-8pt}
\end{figure}

We next adopt a local perspective to better understand the streamwise evolution of these components.
To this end, figure~\ref{fig:Lspodx_PSD} plots local integral-$y^*$-$z^*$-plane power spectral densities for each streamwise station, colored by the \textit{local} SPOD energy ratio \(\uplambda^{(1)}(x)/\sum_i \uplambda^{(i)}(x)\) to quantify the mode dominance: High values (blue) denote energy focused in one leading mode, while low values (red) reflect energy spread across many modes. 
This spatially resolved analysis reveals the statistics and dynamics that dominate at each location and frequency, guiding a subsequent search for coherent structures.
The fundamental spectral evolution starts with the TS wave, marked in figure~\ref{fig:Lspodx_PSD}(a), whose energy level we use as reference for other components (above TS: significant $/$ below TS: weak).
Its FHR $\tilde{\q}$ manifests as dominant peaks that grow with increasing $Re_x$, while non-harmonic components remain negligible in the weak background spectrum. The initial flow is thus deterministic.
As the flow gains energy towards \(C_f^{\text{max}}\) (at $Re_x \approx 4 \times 10^5$), the non-harmonic spectrum gains amplitude with decreasing mode dominance, indicating increasing stochasticity, while the harmonic peaks lose both amplitude and their initial mode dominance. 
Since the flow is cyclo-stationary and all its energy is initially confined to $\tilde{\q}$ (see figure~\ref{fig:STPOD_energy}c), the emergence of broadband content reflects energy transfer from the harmonic components to $\q''^S$, marking the transition from periodic to time-symmetry-broken states. Beyond \(C_f^{\text{max}}\), the spectrum converges towards a fully broadband, low modal dominance state, indicating statistical stationarity and turbulence.

Remarkably, figure~\ref{fig:Lspodx_PSD}(b) shows that even after removing the deterministic FHR $\tilde{\q}$, the \(\q''^S\) fluctuation spectrum still exhibits prominent harmonic peaks with high modal dominance.
This crucial finding indicates that fluctuations remain largely phase-locked to the harmonics initially and can still be represented by few dominant modes at these frequencies, thereby motivating a targeted search for coherent structures responsible for symmetry breaking.

\vspace{-8pt}
\section{The Dynamics and Statistics of Temporal \& Spatial Symmetry Breaking} \label{sec:4}

We apply STPOD to the $\q''^S$ and $\q''^A$ fluctuations using the method outlined in \S~\ref{subsec:STPOD} to identify dominant symmetry breaking modes with minimal assumptions. As we center the data around the FHR $\tilde{\q}$ (which would otherwise become the first mode if not subtracted), we can declare $\bmphi_0 \equiv \tilde{\q}$ as the fundamental space-time mode, leading to the full decomposition
\begin{equation}\label{eq:STPOD_decomp}
\q(\x,t) = \q''(\x,t) + \tilde{\q}(\x, t)
= \sum_{m} \big( \bmphi_m^{S}(\x,\tau) a_{m,j}^{S} + \bmphi_m^{A}(\x,\tau) a_{m,j}^{A} \big) + \bmphi_0(\x,t),
\end{equation}
where $a_{m,j}^{S/A}$ with $j\in [1,N_{T_1}]$ are realization-dependent expansion coefficients describing amplitude modulations of mode $\bmphi_m^{S/A}$ across different flow realizations (periods) $j$.
To further resolve the time dynamics of these space-time modes, we propose a modal multi-timescale approach via a nested space-only POD, separating spatial and temporal information:
\begin{figure}
    \centering
    \includegraphics[trim={0.265cm 0.0cm 0.0cm 0.0cm },clip,width=1.0\textwidth]{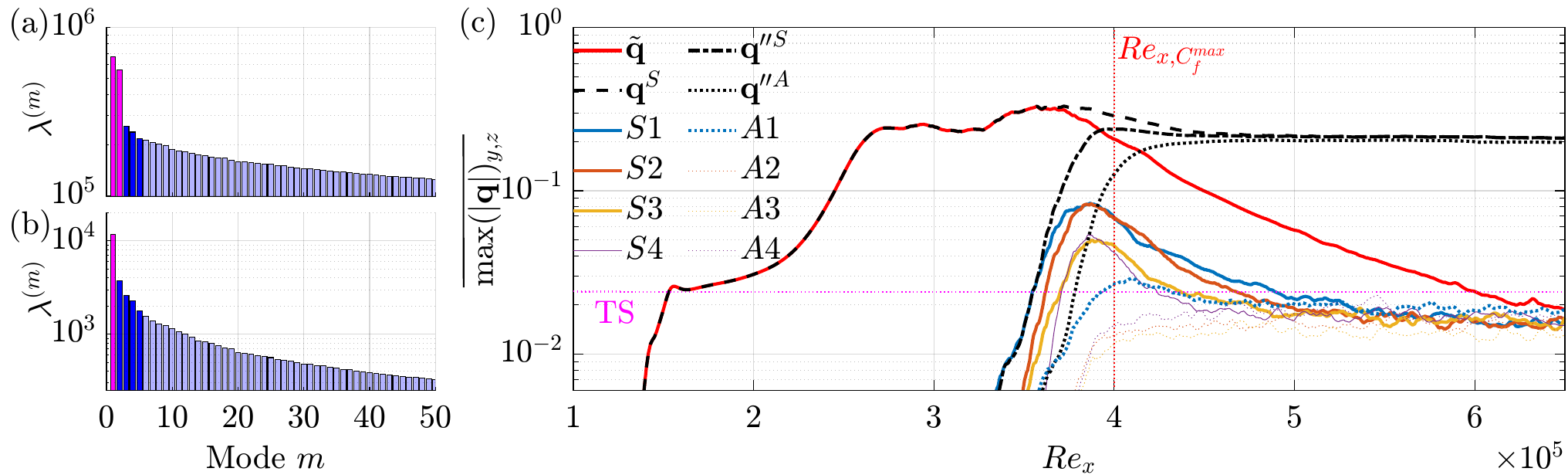}
    \vspace{-15pt}
  \caption{(a) $\q''^S$ symmetric and (b) $\q''^A$ anti-symmetric STPOD modal energy spectra; dominant modes with periodic dynamics colored magenta. (c) Streamwise amplitude development of select STPOD modes $\bmphi_m(\x,\tau) a_{m,j}$, enveloped by the total $\q^S$, $\q^{\prime \prime S}$  and $\q^{\prime \prime A}$ components. The TS wave input amplitude level serves as significance threshold. } 
  \label{fig:STPOD_energy}
  \vspace{-4pt}
\end{figure}
\vspace{-8pt}
\begin{align} \label{eq:POD_STPOD_1}
\bmphi_m(\x,\uptau) &
= \U_m(\x) \textbf{a}_m^{\POD}(\uptau) \\
\q(\x,t) &= \sum_{m} \U_m(\x) \textbf{a}_m^{\POD}(\uptau) a_{m,j} + \U_0(\x) \textbf{a}_0^{\POD}(t). \label{eq:POD_STPOD_2}
\end{align}
We choose POD for its $L_2$-optimality and objectivity about time dynamics, unlike DMD or Fourier methods, which assume linear or periodic dynamics. This decomposition now allows for a clear separation of timescales: the fast local dynamics $\textbf{a}_m^{\POD}(\uptau)$ within a realization ($\uptau \in [0,T_1]$) and slow global modulation $a_{m,j}$ across realizations, where the instantaneous time becomes $t = \uptau+(j-1)T_1$. Furthermore, this enables a dynamical systems analysis by resolving mode trajectory shapes in phase space with $\textbf{a}_m^{\POD}(\uptau)$ \citep{cvitanovic2010geometry}. For visualization, plotting the dominant three coefficients of the $\textbf{a}_m^{\POD} (\uptau)$ vector typically suffices, as higher coefficients generally remain periodic even when dominant ones are not.

\begin{figure}
\centering
\begin{subfigure}[]{1.0\textwidth}
\includegraphics[width=\textwidth]{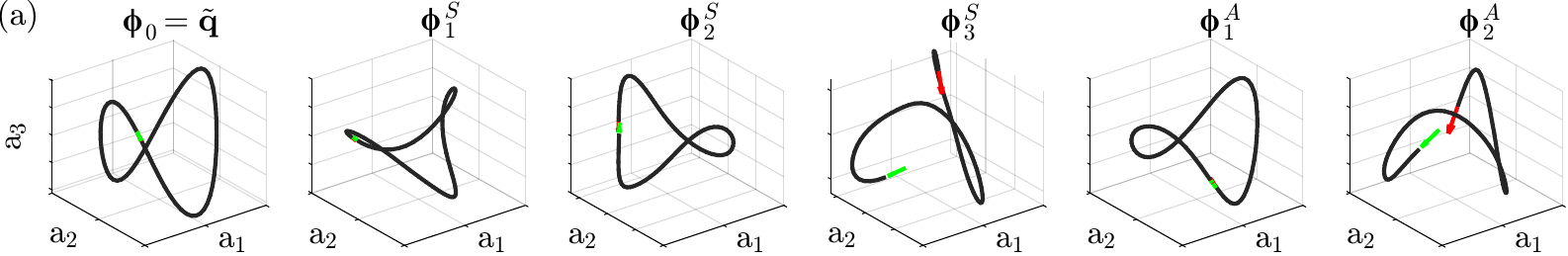}
\end{subfigure}


\begin{subfigure}[]{1.0\textwidth}
\includegraphics[trim={0.0cm 0.0cm 0.0cm 0.0cm },clip,width=\textwidth]{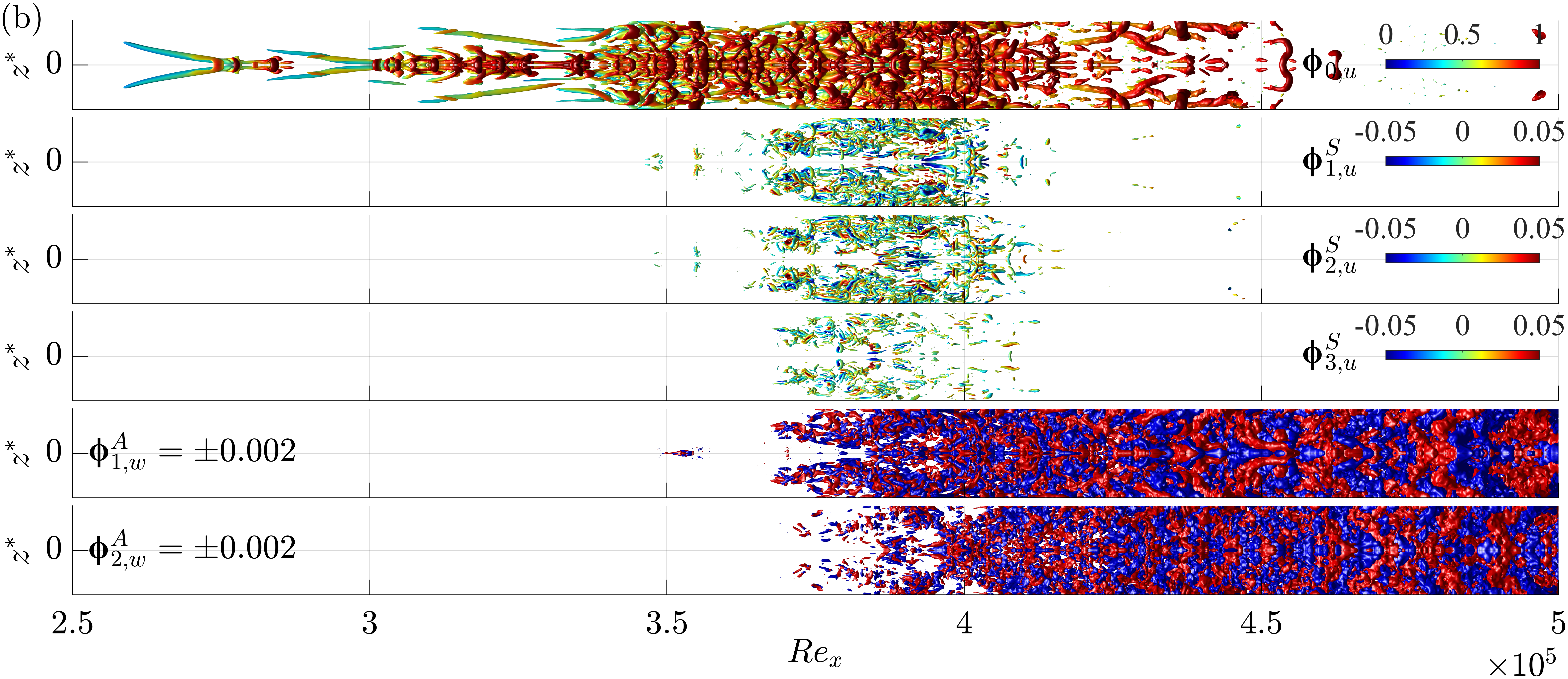}
\end{subfigure}

\begin{subfigure}[]{1.0\textwidth}
\includegraphics[width=\textwidth]{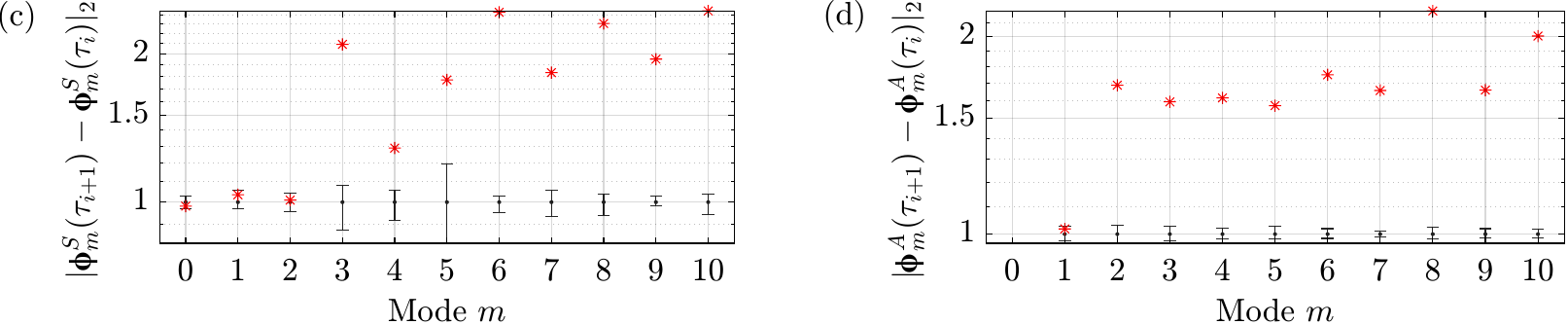}
\end{subfigure}

\vspace{-5pt}    
\caption{Symmetry breaking modes from STPOD. Deterministic mode $\bmphi_0$ ($\equiv$ FHR $\tilde{\q}$), first three symmetric modes ($\bmphi^S_{1-3}$) and first two anti-sym. modes ($\bmphi^A_{1-2}$).
(a) $\textbf{a}_m^{\POD}(\uptau)$ trajectory phase-spaces and 
(b) instantaneous isosurfaces at same phase. Symmetric modes shown by $Q$-criterion ($Q=10^3$) colored by $u$-velocity. Anti-sym. modes shown by $w$-velocity ($w=\pm0.002$).
(c) Symmetric and (d) anti-sym. mode relative $\text{L}_2$-norm distances between consecutive phases: Red markers denote distance from last to first phase (periodicity mismatch), error bars show the existing min$/$max consecutive phase distances within the period, all normalized by the average phase distance in a given mode.}
  \label{fig:STPOD_traj}
  \vspace{-18pt}
\end{figure}

\vspace{-3pt}
\subsection{Onset of Quasi-Periodicity and Aperiodicity}

Figure~\ref{fig:STPOD_energy} plots the energy spectra and streamwise amplitude development of the STPOD modes. Up to $Re_x \lesssim 3.5\times10^5$, the flow is governed entirely by
the deterministic, spatially compact FHR mode of the early transition.
Figure~\ref{fig:STPOD_energy}(c) shows that, as
$\bmphi_0 \equiv\tilde{\q}$ 
falls below the total 
amplitude, 
new 
dominant 
symmetric STPOD modes $\bmphi_1^S$ and $\bmphi_2^S$ surge above the TS wave level around $Re_x \approx 3.5\times10^5$. 
They too are vortical and compact, seen in figure~\ref{fig:STPOD_traj}(b), and
mark the onset of variance from the purely periodic mode $\bmphi_0$. The phase space analysis in figure~\ref{fig:STPOD_traj}(a,c) reveals that these modes, remarkably, also exhibit $T_1$-periodic dynamics:
\begin{equation}
    \bmphi_m^S \left( \mathbf{x}, \uptau \rightarrow T_1 \right) 
    \begin{cases} 
        = \bmphi_m^S  \left( \mathbf{x}, \uptau \rightarrow 0 \right), & m=\{1,2\} \\
        \neq \bmphi_m^S \left( \mathbf{x}, \uptau \rightarrow 0 \right), & m \geq 3.
    \end{cases}
\end{equation}
The critical distinction lies in their non-constant expansion coefficients $a_{1,j}^S$ and $a_{2,j}^S$ that vary over long timescales. Dynamically, this creates a region of \textit{geometric quasi-periodicity} where the state is modulated by different periodic trajectories, i.e. for $3.4\times10^5 \lesssim Re_x \lesssim 3.7\times10^5$,
\begin{equation} \label{eq:STPOD_quasi}
\q(\x,t) \approx \q_{m\le 2}^S(\x,t) = \bmphi_0(\x,t) + a_{1,j}^S\bmphi_1^S(\x,\uptau)  +  a_{2,j}^S \bmphi_2^S(\x,\uptau).
\end{equation}
These trajectories are statistically periodic but instantaneously depart from the deterministic state 
$\bmphi_0$ 
in shapes and directions given by $\bmphi_1^S$ and $\bmphi_2^S$, with 
\textit{amplitudes} and \textit{timings} governed explicitly by their slowly varying expansion coefficients $a_{1,j}^S$ and $a_{2,j}^S$ 
across periods $j$. This quasi-periodic dynamic is a categorical departure from the previous deterministic behavior, as the system now explores multiple periodic trajectories in a structured manner.

\noindent The onset of chaos begins with the emergence of $\bmphi_3^S$ around $Re_x \gtrsim 3.7\times10^5$. This mode breaks the pattern of local periodicity, exhibiting a clear jump
between its first and last phase (figure~\ref{fig:STPOD_traj}a,c). All 
higher
modes ($m \geq 3$) are also aperiodic and located increasingly further downstream. 
This leads to a spectral redistribution from harmonic to broadband frequencies, shown in figure~\ref{fig:Lspodx_PSD}(c): A zoom into the region of critical fluctuation growth reveals the spectral contents of the modes.
Removing the periodic modes $\bmphi_1^S$ and $\bmphi_2^S$ eliminates precisely the mode-dominant harmonic peaks, while removing 
higher
modes flattens the spectrum across all frequencies toward fully broadband characteristics. Since modes that are not $T_1$-periodic can, by definition, \textit{not} be captured solely within the harmonic frequencies, they \textit{must} fill the broadband spectrum between peaks, and are thus indicative of chaos and turbulence.

The progression from periodic to chaotic dynamics seen here echoes the dynamical systems view of wall turbulence \citep{cvitanovic2010geometry, viswanath2007recurrent, kawahara2001periodic}, where trajectories recurrently visit the available unstable periodic orbits that ``scaffold" the state space. Our hierarchical breakdown—from a single periodic base state ($\bmphi_0$), through quasi-periodic modulations ($\bmphi_1^S$, $\bmphi_2^S$), to chaotic excursions via non-periodic modes ($\bmphi_3^S$ and higher)—offers a data-driven perspective on this mechanism.


\subsection{Onset of Spanwise Asymmetry} 
Spatial symmetry breaking follows a remarkably similar route. The anti-symmetric compo-nent $\q^A = \q''^A$ remains energetically negligible until $Re_x\lesssim3.8\times10^5$ (figure~\ref{fig:STPOD_energy}c), from where it rapidly amplifies toward the amplitude of $\q^S$.
To isolate the modes that initiate asymmetry onset, we compute the STPOD for $\q^A$ with a spatial weight focused on the amplified region of interest ($Re_x \leq 4.0\times10^5$) 
prior to the onset of
saturated downstream turbulence. Most notably, the first anti-symmetric mode $\bmphi_1^A$ also exhibits local periodicity (figure~\ref{fig:STPOD_traj}a,d)
\begin{equation}
    \bmphi_m^A \left( \mathbf{x}, \uptau \rightarrow T_1 \right) 
    \begin{cases} 
        = \bmphi_m^A  \left( \mathbf{x}, \uptau \rightarrow 0 \right), & m = 1 \\
        \neq \bmphi_m^A \left( \mathbf{x}, \uptau \rightarrow 0 \right), & m \geq 2
    \end{cases}
\end{equation}
despite a total absence of anti-symmetric forcing or mean flow. 
Similar to equation (\ref{eq:STPOD_quasi}), $\bmphi_1^A$ forms quasi-periodic asymmetric variations with $\bmphi_0$, while spectral analysis analogous to figure~\ref{fig:Lspodx_PSD}(b,c) (omitted for brevity) confirms that higher modes again become broadband.
Thus,
a central finding here is that both spatial and temporal symmetry breaking emerge through coherent space-time structures in distinct dynamical hierarchies, not as random fluctuations.



\vspace{-13.5pt}
\section{Concluding Remarks} 
\vspace{-1pt}
This work reveals the temporal and spatial symmetry breaking mechanisms in canonical K-type boundary layer transition. Deterministic transition begins with 
eigen-modal dynamics 
at a single frequency (TS wave) that evolve into harmonics through quadratic nonlinear interactions. This fundamental harmonic response (FHR) $\tilde{\q}$ is composed entirely of symmetric and periodic coherent structures, is spatially compact and persists far downstream. While $\tilde{\q}$ may resemble turbulence, it remains fully harmonic, clearly defining the extent of the deterministic regime. The ability to pinpoint where deterministic dynamics transition to symmetry breaking provides new clarity in distinguishing organized transition from turbulence onset.

We identify the specific space-time structures responsible for this transition. The FHR $\tilde{\q}$ 
dominates until $Re_x \approx 3.5\times10^5$, after which organized symmetry breaking structures emerge. The dominant two symmetric modes ($\bmphi_1^S$, $\bmphi_2^S$) exhibit periodic dynamics with slow amplitude modulation, thereby deviating from the periodic base state in the form of quasi-periodic trajectories and creating variance around the harmonic peaks. With increasing $Re_x$, higher modes ($\bmphi_3^S$ and beyond) break periodicity and fill the broadband spectrum, delineating the transition to chaos and turbulence.
Notably, spatial symmetry breaking follows a similar organized pattern: despite no anti-symmetric forcings, the first anti-symmetric mode ($\bmphi_1^A$) exhibits periodic dynamics, while higher modes are aperiodic and broadband. 
This shows that both types of symmetry breaking unfold
\textit{not} as random unstructured fluctuations but 
through a hierarchy of emergent, energetically dominant space-time structures, challenging traditional views of symmetry breaking being a purely random process.

Crucially, these findings allow us to define both the onset and spatial reach of new regimes, offering new criteria for quantifying, predicting, and potentially controlling transition. The hierarchical breakdown from a single periodic state ($\bmphi_0\equiv\tilde{\q}$) to quasi-periodic modulation ($\bmphi_1^S$, $\bmphi_2^S$, $\bmphi_1^A$), and finally to increasing degrees of chaos through aperiodic modes, provides new insight into the structure and dynamics governing canonical laminar-turbulent transition.

\backsection[Supplementary data]{\label{SupMat}Supplementary material and movies are available at... } 


\backsection[Funding]{We gratefully acknowledge the NSF for funding this research under grant CBET 2046311.}

\backsection[Declaration of interests]{The authors report no conflict of interest. }




\vspace{-9pt}
\bibliographystyle{jfm}
\bibliography{jfm} 

\begin{thebibliography}{32}
\expandafter\ifx\csname natexlab\endcsname\relax\def\natexlab#1{#1}\fi
\def\au#1{#1} \def\ed#1{#1} \def\yr#1{#1}\def\at#1{#1}\def\jt#1{\textit{#1}} \def\bt#1{#1}\def\bvol#1{\textbf{#1}} \def\vol#1{#1} \def\pg#1{#1} \def\publ#1{#1}\def\arxiv#1{#1}\def\org#1{#1}\def\st#1{\textit{#1}}

\bibitem[Adrian(2007)]{adrian2007hairpin}
{\sc \au{Adrian, R.~J.}} \yr{2007}  \at{Hairpin vortex organization in wall turbulence}.  \jt{Phys. Fluids}  \bvol{19}.

\bibitem[Aubry(1991)]{aubry1991hidden}
{\sc \au{Aubry, N.}} \yr{1991}  \at{{On the hidden beauty of the proper orthogonal decomposition}}.  \jt{Theor. Comput. Fluid Dyn.}  \bvol{2},  \pg{339--352}.

\bibitem[Bake {\em et~al.\/}(2002)Bake, Meyer \& Rist]{bake2002turbulence}
{\sc \au{Bake, S.}, \au{Meyer, D. G.~W.} \& \au{Rist, U.}} \yr{2002}  \at{{Turbulence mechanism in Klebanoff transition: a quantitative comparison of experiment and direct numerical simulation}}.  \jt{J. Fluid Mech.}  \bvol{459},  \pg{217--243}.

\bibitem[Cherubini {\em et~al.\/}(2011)Cherubini, De~Palma, Robinet \& Bottaro]{cherubini2011minimal}
{\sc \au{Cherubini, S.}, \au{De~Palma, P.}, \au{Robinet, J.-C.} \& \au{Bottaro, A.}} \yr{2011}  \at{{The minimal seed of turbulent transition in the boundary layer}}.  \jt{J. Fluid Mech.}  \bvol{689},  \pg{221--253}.

\bibitem[Cvitanovi{\'c} \& Gibson(2010)]{cvitanovic2010geometry}
{\sc \au{Cvitanovi{\'c}, P.} \& \au{Gibson, J.~F.}} \yr{2010}  \at{Geometry of the turbulence in wall-bounded shear flows: periodic orbits}.  \jt{Phys. Scr.}  \bvol{2010},  \pg{014007}.

\bibitem[Fasel {\em et~al.\/}(1990)Fasel, Rist \& Konzelmann]{fasel1990numerical}
{\sc \au{Fasel, H.~F.}, \au{Rist, U.} \& \au{Konzelmann, U.}} \yr{1990}  \at{{Numerical investigation of the three-dimensional development in boundary-layer transition}}.  \jt{AIAA J.}  \bvol{28},  \pg{29--37}.

\bibitem[Frame \& Towne(2023)]{frame2023space}
{\sc \au{Frame, P.} \& \au{Towne, A.}} \yr{2023}  \at{{Space-time {POD} and the Hankel matrix}}.  \jt{PLoS One}  \bvol{18},  \pg{e0289637}.

\bibitem[Hack \& Schmidt(2021)]{hack2021extreme}
{\sc \au{Hack, M. J.~P.} \& \au{Schmidt, O.~T.}} \yr{2021}  \at{{Extreme events in wall turbulence}}.  \jt{J. Fluid Mech.}  \bvol{907},  \pg{A9}.

\bibitem[Heidt \& Colonius(2024)]{heidt2024spectral}
{\sc \au{Heidt, L.} \& \au{Colonius, T.}} \yr{2024}  \at{{Spectral proper orthogonal decomposition of harmonically forced turbulent flows}}.  \jt{J. Fluid Mech.}  \bvol{985},  \pg{A42}.

\bibitem[Herbert(1988)]{herbert1988secondary}
{\sc \au{Herbert, T.}} \yr{1988}  \at{{Secondary instability of boundary layers}}.  \jt{Annu. Rev. Fluid Mech.}  \bvol{20},  \pg{487--526}.

\bibitem[Kachanov {\em et~al.\/}(1977)Kachanov, Kozlov \& Levchenko]{kachanov1977nonlinear}
{\sc \au{Kachanov, Y.~S.}, \au{Kozlov, V.~V.} \& \au{Levchenko, V.~Y.}} \yr{1977}  \at{{Nonlinear development of a wave in a boundary layer}}.  \jt{Fluid Dyn.}  \bvol{12},  \pg{383--390}.

\bibitem[Kachanov \& Levchenko(1984)]{kachanov1984resonant}
{\sc \au{Kachanov, Y.~S.} \& \au{Levchenko, V.~Y.}} \yr{1984}  \at{{The resonant interaction of disturbances at laminar-turbulent transition in a boundary layer}}.  \jt{J. Fluid Mech.}  \bvol{138},  \pg{209--247}.

\bibitem[Kawahara \& Kida(2001)]{kawahara2001periodic}
{\sc \au{Kawahara, G.} \& \au{Kida, S.}} \yr{2001}  \at{Periodic motion embedded in plane couette turbulence: regeneration cycle and burst}.  \jt{J. Fluid Mech.}  \bvol{449},  \pg{291--300}.

\bibitem[Klebanoff {\em et~al.\/}(1962)Klebanoff, Tidstrom \& Sargent]{klebanoff1962three}
{\sc \au{Klebanoff, P.~S.}, \au{Tidstrom, K.~D.} \& \au{Sargent, L.~M.}} \yr{1962}  \at{{The three-dimensional nature of boundary-layer instability}}.  \jt{J. Fluid Mech.}  \bvol{12},  \pg{1--34}.

\bibitem[Lin \& Schmidt(2024)]{lin2024modal}
{\sc \au{Lin, C.} \& \au{Schmidt, O.~T.}} \yr{2024}  \bt{{Modal decomposition of {K}-type boundary layer transition}}.  \pg{p. 0495}.  \publ{{AIAA SCITECH 2024 FORUM}}.

\bibitem[Lumley(1970)]{Lumley1970AP}
{\sc \au{Lumley, J.~L.}} \yr{1970} {\em {Stochastic Tools in Turbulence}\/}, 1st edn.  \publ{Academic Press}.

\bibitem[Monokrousos {\em et~al.\/}(2010)Monokrousos, Akervik, Brandt \& Henningson]{monokrousos2010global}
{\sc \au{Monokrousos, A.}, \au{Akervik, E.}, \au{Brandt, L.} \& \au{Henningson, D.~S.}} \yr{2010}  \at{{Global three-dimensional optimal disturbances in the Blasius boundary-layer flow using time-steppers}}.  \jt{J. Fluid Mech.}  \bvol{650},  \pg{181--214}.

\bibitem[Rempfer \& Fasel(1994{\natexlab{{\em a\/}}})]{rempfer1994dynamics}
{\sc \au{Rempfer, D.} \& \au{Fasel, H.~F.}} \yr{1994{\natexlab{{\em a\/}}}}  \at{{Dynamics of three-dimensional coherent structures in a flat-plate boundary layer}}.  \jt{J. Fluid Mech.}  \bvol{275},  \pg{257--283}.

\bibitem[Rempfer \& Fasel(1994{\natexlab{{\em b\/}}})]{rempfer1994evolution}
{\sc \au{Rempfer, D.} \& \au{Fasel, H.~F.}} \yr{1994{\natexlab{{\em b\/}}}}  \at{{Evolution of three-dimensional coherent structures in a flat-plate boundary layer}}.  \jt{J. Fluid Mech.}  \bvol{260},  \pg{351--375}.

\bibitem[Rigas {\em et~al.\/}(2021)Rigas, Sipp \& Colonius]{rigas2021nonlinear}
{\sc \au{Rigas, G.}, \au{Sipp, D.} \& \au{Colonius, T.}} \yr{2021}  \at{{Nonlinear input/output analysis: application to boundary layer transition}}.  \jt{J. Fluid Mech.}  \bvol{911},  \pg{A13}.

\bibitem[Rist \& Fasel(1995)]{rist1995direct}
{\sc \au{Rist, U.} \& \au{Fasel, H.~F.}} \yr{1995}  \at{{Direct numerical simulation of controlled transition in a flat-plate boundary layer}}.  \jt{J. Fluid Mech.}  \bvol{298},  \pg{211--248}.

\bibitem[Sayadi {\em et~al.\/}(2013)Sayadi, Hamman \& Moin]{sayadi2013direct}
{\sc \au{Sayadi, T.}, \au{Hamman, C.~W.} \& \au{Moin, P.}} \yr{2013}  \at{{Direct numerical simulation of complete H-type and K-type transitions with implications for the dynamics of turbulent boundary layers}}.  \jt{J. Fluid Mech.}  \bvol{724},  \pg{480}.

\bibitem[Schmidt \& Colonius(2020)]{SchmidtColonius2020AIAAJ}
{\sc \au{Schmidt, O.~T.} \& \au{Colonius, T.}} \yr{2020}  \at{{Guide to Spectral Proper Orthogonal Decomposition}}.  \jt{AIAA J.}  \bvol{58},  \pg{1023--1033}.

\bibitem[Schmidt \& Schmid(2019)]{schmidt2019conditional}
{\sc \au{Schmidt, O.~T.} \& \au{Schmid, P.~J.}} \yr{2019}  \at{{A conditional space--time {POD} formalism for intermittent and rare events: example of acoustic bursts in turbulent jets}}.  \jt{J. Fluid Mech.}  \bvol{867},  \pg{R2}.

\bibitem[Sirovich(1987{\natexlab{{\em a\/}}})]{sirovich1987turbulence_1}
{\sc \au{Sirovich, L.}} \yr{1987{\natexlab{{\em a\/}}}}  \at{{Turbulence and the dynamics of coherent structures. I. Coherent structures}}.  \jt{Q. Appl. Math.}  \bvol{45},  \pg{561--571}.

\bibitem[Sirovich(1987{\natexlab{{\em b\/}}})]{sirovich1987turbulence_2}
{\sc \au{Sirovich, L.}} \yr{1987{\natexlab{{\em b\/}}}}  \at{{Turbulence and the dynamics of coherent structures. II. Symmetries and transformations}}.  \jt{Q. Appl. Math.}  \bvol{45},  \pg{573--582}.

\bibitem[Towne {\em et~al.\/}(2018)Towne, Schmidt \& Colonius]{TowneEtAl2018JFM}
{\sc \au{Towne, A.}, \au{Schmidt, O.~T.} \& \au{Colonius, T.}} \yr{2018}  \at{{Spectral proper orthogonal decomposition and its relationship to dynamic mode decomposition and resolvent analysis}}.  \jt{J. Fluid Mech.}  \bvol{847},  \pg{821--867}.

\bibitem[Tutkun \& George(2017)]{tutkun2017lumley}
{\sc \au{Tutkun, M.} \& \au{George, W.~K.}} \yr{2017}  \at{{Lumley decomposition of turbulent boundary layer at high Reynolds numbers}}.  \jt{Phys. Fluids}  \bvol{29},  \pg{025108}.

\bibitem[Viswanath(2007)]{viswanath2007recurrent}
{\sc \au{Viswanath, D.}} \yr{2007}  \at{Recurrent motions within plane couette turbulence}.  \jt{J. Fluid Mech.}  \bvol{580},  \pg{339--358}.

\bibitem[Welch(1967)]{Welch1967IEEE}
{\sc \au{Welch, P.~D.}} \yr{1967}  \at{The use of fast fourier transform for the estimation of power spectra: A method based on time averaging over short, modified periodograms}.  \jt{IEEE Trans. Audio Electroacoust.}  \bvol{15},  \pg{70--73}.

\bibitem[White(2006)]{white2006viscous}
{\sc \au{White, F.~M.}} \yr{2006} {\em {Viscous fluid flow}\/}, 3rd edn.  \publ{McGraw-Hill}.

\bibitem[Wu \& Moin(2009)]{wu2009direct}
{\sc \au{Wu, X.} \& \au{Moin, P.}} \yr{2009}  \at{{Direct numerical simulation of turbulence in a nominally zero-pressure-gradient flat-plate boundary layer}}.  \jt{J. Fluid Mech.}  \bvol{630},  \pg{5--41}.

\end{thebibliography}


\end{document}